# Topological one-way Weyl fiber


Hao Lin[1,*], Yu Wang[1,*], Zitao Ji[1,*], Yidong Zheng[1], Jianfeng Chen[1,2,†] and Zhi-Yuan Li[1,3,†]

[1] School of Physics and Optoelectronics, South China University of Technology, Guangzhou 510640, China

[2] Department of Electrical and Computer Engineering, National University of Singapore, Singapore 117583, Singapore

[3] State Key Laboratory of Luminescent Materials and Devices, South China University of Technology, Guangzhou, 510640, China

[*] These authors contributed equally to this work.

[†] Correspondence: phzyli@scut.edu.cn; jfchen@nus.edu.sg



**Abstract:** Topological photonics enables unprecedented photon manipulation by realizing various topological states, such as corner states, edge states, and surface states. However, achieving a topological fiber state has remained elusive. Here, we demonstrate a topological fiber state in a Weyl gyromagnetic photonic crystal fiber. By applying an in-plane magnetic bias to a gyromagnetic photonic crystal fiber with broken parity-inversion symmetry, we create an asymmetrical Weyl bandgap that supports one-way fiber states associated with type-II Weyl points. Dispersion and topological invariant calculations reveal the transition from Weyl surface states to one-way Weyl fiber states. Electromagnetic field simulations confirm the existence of one-way Weyl fiber states and their robust transport in the presence of metallic obstacle along the transport path. Our findings offer an intriguing pathway for exploring novel topological states and guiding the design of topological fibers.




**Introduction.** Topological photonics has sparked significant interest due to phenomena associated with the robust transport of light and electromagnetic waves [1-4]. Unlike trivial photonic states, topological states are guaranteed by bulk-edge correspondence, where bulk topological invariants determine their existence [5,6]. Hallmark photonic topological phenomena, such as unidirectional backscattering-immune photonic edge/surface states, have been discovered in two-dimensional (2D) and three-dimensional (3D) photonic crystal with broken time-reversal symmetry [2-4,7-13]. Additionally, a variety of photonic topological states, including spin and valley Hall edge states [14-16], higher-order corner and hinge states [17-21], and surface states [22,23], have been observed. These topological states are protected against deformations or symmetry alterations, making them ideal for applications in integrated [24,25], nonlinear [26,27], and quantum photonics [28,29]. However, realizing a topological fiber state has remained elusive. Fiber states are crucial for the communication and interconnection of light and electromagnetic waves [30-33].

In this Letter, we demonstrate a Weyl gyromagnetic photonic crystal fiber that hosts a topological one-way fiber state within an asymmetrical Weyl bandgap. Conventional photonic crystal fibers typically support reciprocal fiber states, where any defects lead to significant backscattering [Fig. 1(a)]. In contrast, the Weyl photonic crystal fiber proposed here supports a topologically protected one-way fiber state associated with type-II Weyl points, which is immune to imperfections and obstacles along the transport path [Fig. 1(b)]. This is achieved by applying an in-plane magnetic bias to a gyromagnetic photonic crystal fiber with broken inversion symmetry, leading to the emergence of Weyl points and nonreciprocity along the transport direction. We showcase the process of using Weyl surface states to construct one-way Weyl fiber states, directly illustrating their intrinsic connection. Additionally, we demonstrate the unique properties of these topological fiber states, including one-way robust propagation, tilted Weyl dispersion, and asymmetrical Weyl bandgap.



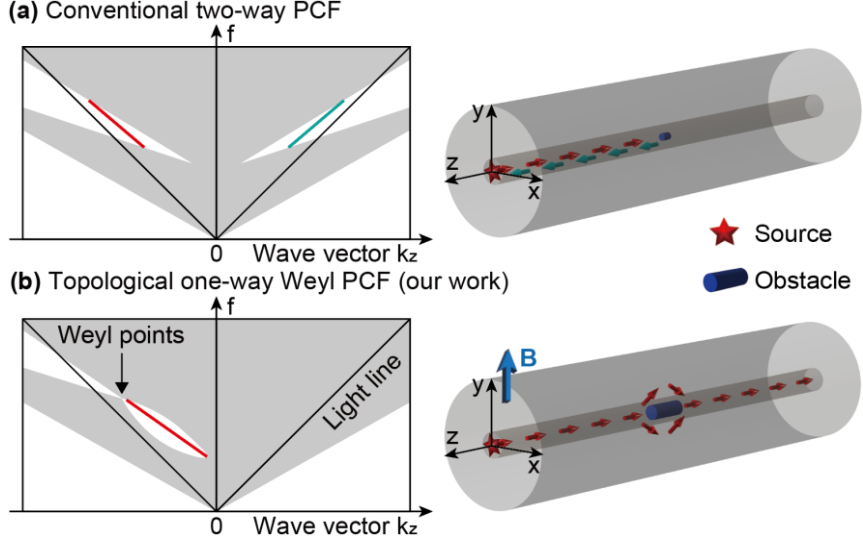

**FIG. 1. Conventional and Weyl photonic crystal fibers.** (a) Conventional photonic crystal fiber (PCF). Significant backscattering occurs when light and electromagnetic waves encounter a small defect or obstacle within the fiber. (b) Weyl photonic crystal fiber with an in-plane magnetic bias (blue arrow). Weyl fiber states propagate unidirectionally and are immune to big obstacles. The red and cyan bands indicate fiber states transport along the $-z$ (red arrows) and $+z$ (cyan arrows) directions, respectively.

**Dispersions and topological characteristics.** We initiate our exploration with a gyromagnetic photonic crystal featuring a rectangular lattice, as depicted in Fig. 2(a). The material properties of gyromagnetic materials (yttrium iron garnet, YIG) can be found in Supplemental Materials. The lattice constants along the $x$ and $y$ directions are $a$=11 mm and $b$=7.7 mm, respectively. The unit cell of this lattice consists of three YIG rods ($\varepsilon$=14.5) with distinct diameters ($d_1$, $d_1$, $d_2$) positioned in air. The distance between their centers forms an equilateral triangle with a length of $R$=3.9 mm. Consequently, the parity-inversion symmetry of the photonic crystal is broken. Furthermore, the photonic crystal exhibits continuous translational symmetry along the $z$-axis, allowing its eigenmodes to be characterized through a sequence of 2D Brillouin zones (BZs) dependent on the wave vector $k_z$, as shown in Fig. 2(b).



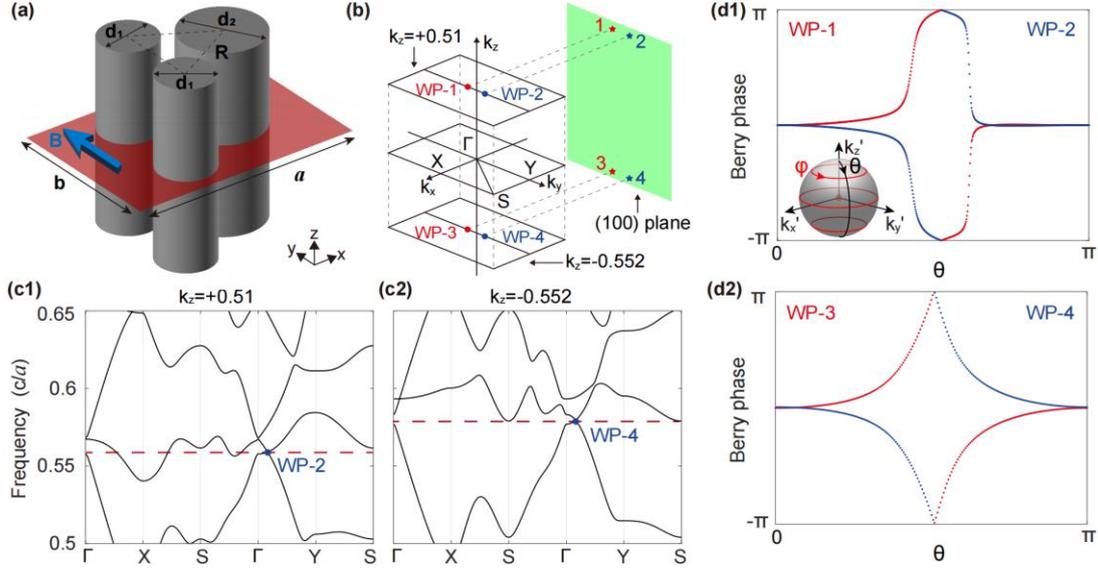

**FIG. 2. Dispersions and topological characteristics.** (a) The unit cell (red plane) includes three YIG rods ($\varepsilon$=14.5) of distinct diameters ($d_1$=2.7 mm, $d_2$=3.9 mm) in air, spaced with a center distance of $R$=3.9 mm. The lattice constants along the $x$ and $y$ directions are $a$=11 mm and $b$=7.7 mm, respectively. The blue arrow indicates the in-plane external magnetic bias. (b) The $k_z$-dependent 2D BZs alongside the projected (100) surface BZ. Two pairs of Weyl points (WP-1&2, WP-3&4) are highlighted, with their projections shown by red and blue stars. (c1, c2) Bulk bands for (c1) $k_z$=+0.51 and (c2) $k_z$=−0.552. The frequencies of the red dotted lines are 0.56 $c/a$ and 0.58 $c/a$, respectively. (d1, d2) Berry phases of Weyl points: (d1) WP-1&2, (d2) WP-3&4. The inset in (d1) illustrates a momentum sphere encompassing a Weyl point, with red closed paths employed for Berry Phase calculations.

To construct a Weyl photonic crystal, we apply an external magnetic bias along the $y$ direction to break the time-reversal symmetry of the system. This operation disrupts all the symmetries ($M_{xy}$, $C_{2x}$, $T$) related to the $z$-axis, enabling the emergence of an asymmetric Weyl bandgap. Figures 2(c1) and 2(c2) illustrate the bulk bands along the specified high-symmetry lines in the $k_z$-dependent BZ (i.e., $k_z$=+0.51 and $k_z$=−0.552). There exist two pairs of type-II Weyl points between the third and fourth photonic bands. They are represented by red (positive



chirality) and blue (negative chirality) dots in the $k_z$-dependent BZ [Fig. 2(b)], with coordinates $(k_x, k_y, k_z) = (0, \pm0.08, +0.51)$ and $(0, \pm0.08, -0.552)$. The units for $k_y$ and $k_z$ are normalized to $2\pi/b$ and $2\pi/a$, respectively.

Each pair of Weyl points can be related through the mirror operation ($M_{xz}$) of the system. The Weyl points at $k_z=-0.552$ are ideal type-II Weyl points (well-separated in frequency) [34], while those at $k_z=+0.51$ are not (See Figs. 2(c1) and 2(c2)). The topological charge (or chirality) of each Weyl point is determined using the Wilson-loop method [35,36]. We select a momentum sphere in $k$-space with a radius of $d_k=0.05$ around a Weyl point, as shown in Fig. 2(d1). For each polar angle $\theta$, we calculate the Berry phase of the third band along a closed loop, varying the azimuthal angle $\phi$ from 0 to $2\pi$. The results, depicted in Figs. 2(d1) and 2(d2), showcase the Berry phase evolution as $\theta$ ranges from 0 to $\pi$, revealing the topological charges ($C=+1, -1$) associated with each Weyl point. We develop an effective Hamiltonian to describe each pair of Type-II Weyl points in the Supplemental Materials.

**Weyl surface states and Weyl fiber states.** To examine the Weyl surface states at the (100) surfaces, we construct a supercell with 11 cycles along the $x$ direction (Supercell-1), as shown in Fig. 3(a). The supercell's left and right boundaries are set as perfect electric conductor (PEC) boundaries, while the top and bottom boundaries have periodic boundary conditions. We select a rectangular yellow path around WP-4 on the (100) surface BZ to discuss the dispersion behavior [Fig. 3(b)]. Along this path, the band is characterized by a well-defined Chern number ($C=-1$), see Supplemental Materials for the detailed calculations. The projected bandstructure shows a single gapless surface state on each (100) surface within the bandgap, colored orange and purple. The corresponding eigenmodal fields-1&2 are shown in Fig. 3(d). We further construct Supercell-2 with an intermediate defect by removing three YIG rods [Fig. 3(a)], which can be seen as the proximity and connection of two Supercell-1 units. Figure 3(c)



displays two newly formed defect modes emerging in the bandgap, due to the proximity coupling of two Weyl surface states from distinct interfaces of the Weyl crystal [37]. Their eigenmodal fields-3&4 can be seen in Fig. 3(d). Such hybrid Weyl surface states maintain stability because they originate from the coupling of robust Weyl surface states.

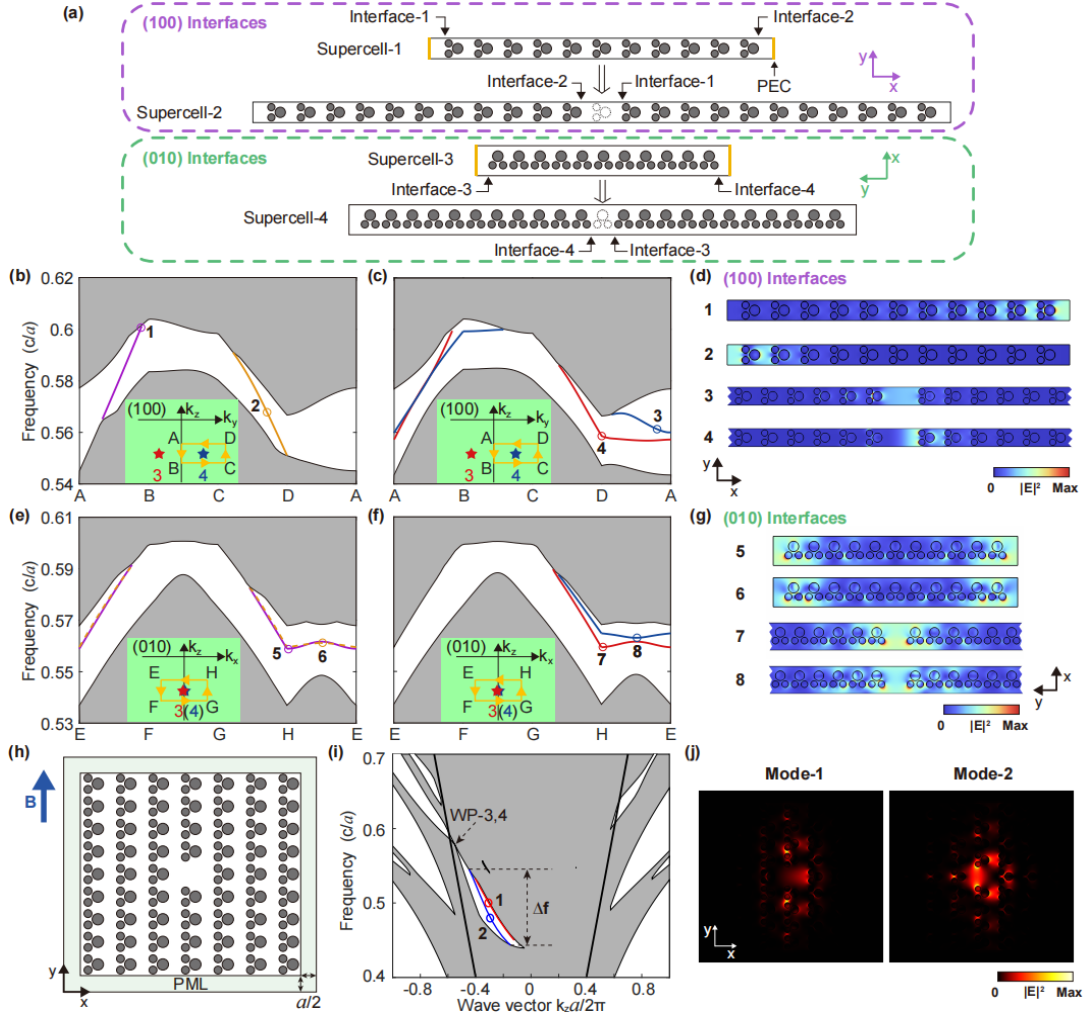

**FIG. 3. Weyl surfaces states and Weyl fiber states.** (a) The left and right boundaries of Supercell-1 (Supercell-3) are PEC boundaries. Supercell-2 (Supercell-4) consists of two Supercell-1 (Supercell-3) units positioned head-to-head along the $x$ ($y$) direction with distances $a$ ($b$). Periodic boundary conditions are applied to the other remaining boundaries. (b) Projected bandstructure of Supercell-1 along a yellow rectangular path (A-B-C-D-A). The purple and orange curves represent the Weyl surface states. (c) Projected bandstructure of Supercell-2. The



red and blue curves represent hybrid Weyl surface states. (d) Eigenmodal fields of Supercell-1&2. (e) Projected bandstructure of Supercell-3 along a yellow rectangular path (E-F-G-H-E). The purple and orange dotted curve represent the trivial surface states. (f) Projected bandstructure of Supercell-4. The red and blue curves represent hybrid surface states. (g) Eigenmodal fields of Supercell-3&4. (h) End-face of a Weyl photonic crystal fiber (7×9 layers). (i) The Weyl fiber's bandstructure with two one-way fiber states (red and blue curves) within an asymmetric bandgap. (j) Eigenmodal fields of Weyl fiber states.

We investigate the properties of the (010) surface states using a similar approach. Initially, we construct an 11-period supercell along the *y*-axis, designated as Supercell-3 [Fig. 3(a)]. On the (010) surface BZ, Weyl points 3&4 are projected onto the same location. A rectangular path encircling this projection is selected, as depicted in Fig. 3(e). Due to the superposition of positive and negative chiralities, the Chern number of the path is zero, indicating trivial (010) surface states. The two surface states (degenerate) and their modal fields are shown in Figs. 3(e) and 3(g), respectively. Unlike the (100) surface states that spiral through the bandgap, the (010) surface states are trivial and easily disturbed by the boundary environment. Similarly, Supercell-4 is constructed [Fig. 3(a)] to demonstrate the coupling of the two (010) surface states. The dispersion and field profiles of the resulting two hybrid surface states are presented in Figs. 3(f) and 3(g). These hybrid surface states are also fragile because they depend on the two trivial (010) surface states of Supercell-3.

We construct a Weyl photonic crystal fiber [Fig. 3(h)], surrounded by perfectly matched layer (PML) cladding (light green area). The fiber's hollow-core defect can be seen as the superposition of the four interfaces of Supercell-1&3. Thus, the fiber modes result from the further interaction and coupling of the intermediate hybrid states of Supercell-2&4. Since the former is robust and the latter is trivial, only the (100) surface states persist and interact to



shape the fiber modes during such boundary construction, particularly for hybrid Weyl surface states predominantly moving along the -$k_z$ direction [Fig. 3(c)]. Figure 3(i) displays the fiber's bandstructure, highlighting the Weyl fiber states within the asymmetric Weyl bandgap with red and blue curves. The fields of these two fiber states are confined within the fiber core [Fig. 3(j)]. The fiber bands are connected to the Weyl points and remains stable within the Weyl bandgap due to the topological protection. The slope of these one-way Weyl fiber states shows their group velocity ($v_g = d\omega/dk$), indicating transport along the -$z$ direction. No fiber modes appear at +$k_z$ due to the absence of a bandgap.

**Identification of one-way Weyl fiber states.** To demonstrate the transport performance of the one-way Weyl fiber states, we construct a Weyl photonic crystal fiber with dimensions of 7$a$ along the $x$-direction and 9$b$ along the $y$-direction, and $h$=80 cm along the $z$-direction [Fig. 4(a)]. A point source (green star) at a frequency of $f$=0.48 $c/a$ is positioned at the center to excite the fiber states. Figure 4(b1) illustrates the one-way transport behavior (downward) of the Weyl fiber states. This is further confirmed by the field distributions at the top and bottom ends [Fig. 4(c1)]. We define a contrast coefficient, $\eta = 10 \lg(P_{bot}/P_{top})$, to quantify the one-way transport, where $P_{bot}$ and $P_{top}$ indicate the output power at the bottom and top ends, respectively. Numerical calculations reveal that $P_{bot} = 9943.8$ W, $P_{top} = 149.7$ W, and $\eta = 18.22$ dB, indicating a considerable contrast between downward and upward transmission. Figure 4(d) shows that a wide frequency range from 0.47 to 0.56 $c/a$ exhibits a contrast coefficient of over 10 dB.

To verify the transport robustness of the one-way Weyl fiber states, we insert a metallic cylinder with radius $r$=0.3$a$ and length $H$=4$a$ into the transport path at $z$=$h$/4. Notably, the length of the metallic cylinder is about twice the excited wavelength, making it a very strong scatterer. Figures 4(b2) and 4(c2) clearly demonstrate that the energy fluxes can bypass the



metallic obstacle and continue downward transport without backscattering ($\eta = 16.48$ dB). Figure 4(e) illustrates the contrast coefficient varying with the length of metallic obstacle at an excitation frequency of $f$=0.48 $c/a$. Even with the obstacle length increasing to approximately five times the excited wavelength ($\approx 2a$), the contrast coefficient remains high (about 14 dB). We discuss the basic parameters of the one-way Weyl fiber states, including confinement loss, group velocity, and group velocity dispersion in the Supplemental Materials.

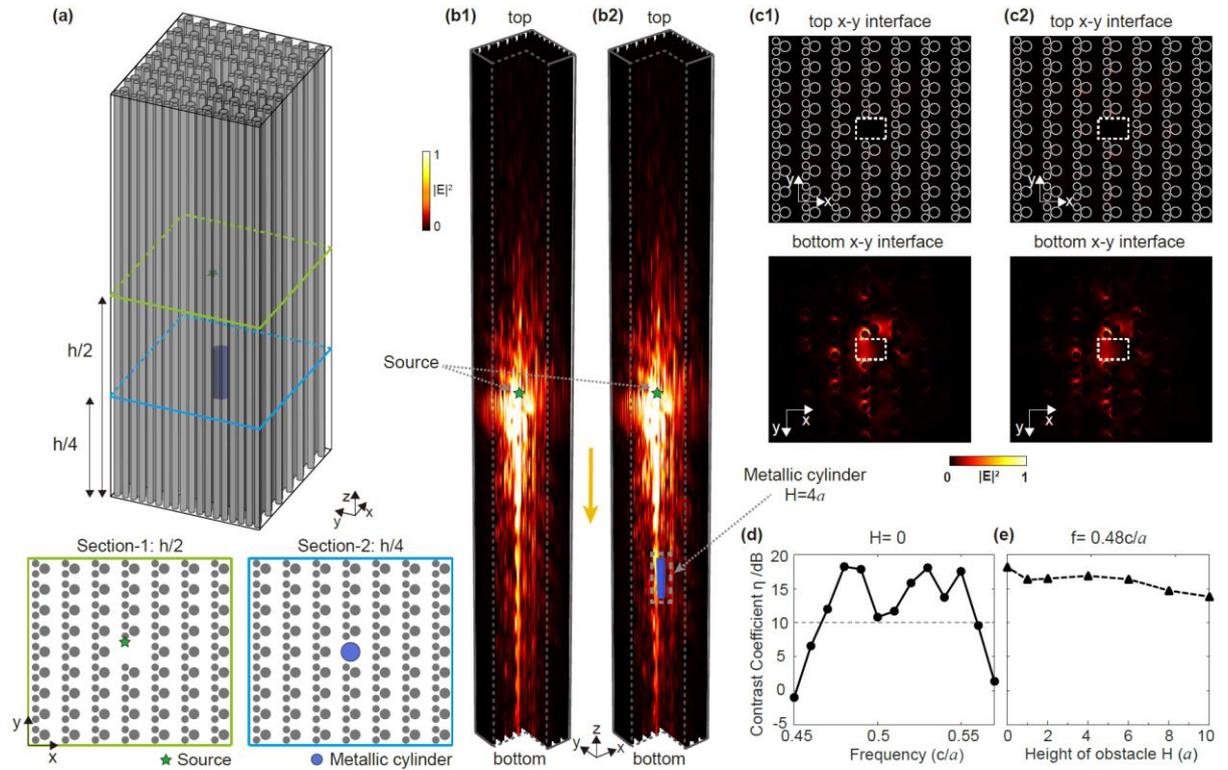

**FIG. 4. Identification of one-way Weyl fiber states.** (a) Schematic of the Weyl photonic crystal fiber. A point source (green star) is positioned at the center. A metallic cylinder (blue) is inserted to verify the robust transport of one-way Weyl fiber states. (b1, b2) Transport of electromagnetic waves in a Weyl photonic crystal fiber (b1) without and (b2) with metallic obstacles. The frequency of the source is $f$=0.48 $c/a$. (c1, c2) Output electric fields at the top and bottom ends in the cases of (c1) without and (c2) with metallic obstacles. (d) Contrast



coefficient across frequencies from 0.45 to 0.57 $c/a$ without metallic obstacle. (e) Contrast

coefficient at $f$=0.48 $c/a$ vs. obstacle height from 0 to 10$a$.

**Conclusion and outlook.** We have demonstrated a Weyl photonic crystal fiber that supports

one-way Weyl fiber states. This is achieved by applying an in-plane external magnetic bias to

a gyromagnetic photonic crystal with broken parity-inversion symmetry, thereby driving the

emergence of one-way fiber states within an asymmetric Weyl bandgap. Dispersions and

eigenmodal fields reveal the transition of Weyl topological states from Weyl surface states to

hybrid Weyl surface states and finally to one-way Weyl fiber states. Both transport behaviors

and transmission spectra confirm the topological nature of the one-way Weyl fiber states, which

remain robust against imperfections and obstacles. We noticed a notable theoretical design

claiming the realization of one-way fiber states in 3D gyromagnetic photonic crystals [38]. It

relies on 3D helical modulation rather than the common fiber configuration, making

implementation challenging. In contrast, our design requires only a simple arrangement of

gyromagnetic cylinders.


**Acknowledgements**

The authors are grateful for the financial support from Guangdong Innovative and

Entrepreneurial Research Team Program (2016ZT06C594), Science and Technology Project of

Guangdong (2020B010190001), National Key R&D Program of China (2018YFA0306200),

and the National Natural Science Foundation of China (11974119).

# Supplemental Materials for

# Topological one-way Weyl fiber


Hao Lin[1,*], Yu Wang[1,*], Zitao Ji[1,*], Yidong Zheng[1], Jianfeng Chen[1,2,†] and Zhi-Yuan Li[1,3,†]

[1] School of Physics and Optoelectronics, South China University of Technology, Guangzhou 510640, China

[2] Department of Electrical and Computer Engineering, National University of Singapore, Singapore 117583, Singapore

[3] State Key Laboratory of Luminescent Materials and Devices, South China University of Technology, Guangzhou, 510640, China

[*] These authors contributed equally to this work.

[†] Correspondence: phzyli@scut.edu.cn; jfchen@nus.edu.sg


**I. Material properties.**

**II. Effective Hamiltonian for Weyl Points.**

**III. The Berry phase evolution along the closed path in the (100) surface BZ.**

**IV. The performance of the one-way fiber modes.**



**I. Material properties.** The material parameters of the YIG rods discussed in the main text are obtained from the commercially available gyromagnetic material yttrium iron garnet (YIG), which possesses a relative permittivity of ε=14.5. In the absence of an external magnetic field, the relative permeability is μ=1. Upon the application of a direct current magnetic field along the $y$ axis of the system, the YIG rods become magnetized. Consequently, the relative permeability tensor of the YIG rods adopts the following form:

$$\hat{\mu} = \begin{pmatrix} \mu_r & 0 & i\mu_k \\ 0 & 1 & 0 \\ -i\mu_k & 0 & \mu_r \end{pmatrix}$$

Where $\mu_r = 1 + \frac{(\omega_0 + i\alpha\omega)\omega_m}{(\omega_0 + i\alpha\omega)^2 - \omega^2}$ , $\mu_k = \frac{\omega\omega_m}{(\omega_0 + i\alpha\omega)^2 - \omega^2}$ , $\omega_m = \gamma M_s$ , $\omega_0 = \gamma H_0$, $\omega = 2\pi f$

represents the operating angular frequency, $\gamma = 1.76 \times 10^{11} \text{s}^{-1}\text{T}^{-1}$ is the gyromagnetic ratio, the saturation magnetization is $M_s = 1950$ Guass, and the damping coefficient α can be simply set as 0. When the external magnetic field is set to $H_0 = 2000$ Oe, and the operating frequency to $f = 14$ GHz in our work, the resulting permeability tensor elements are $\mu_r = 0.814$ and $\mu_k = 0.464$.

**II. Effective Hamiltonian for Weyl Points.** In our symmetry-broken system, the two pair of Weyl points at plus and minus $k_z$ are decoupled. The dispersion of each pair of type-II Weyl points, as discussed in the main text, can be locally described by a $2 \times 2$ Weyl Hamiltonian:

$\boldsymbol{H(k)} = v_x k_x \sigma_x + v_y (k_y^2 - q_y^2) \sigma_y + v_z (k_z - q_z)(I - \gamma \sigma_z)$ . Here, $k_x$, $k_y$ and $k_z$ are the wave vector components; $v_x$, $v_y$ and $v_z$ are the group velocities; $\sigma_x$, $\sigma_y$ and $\sigma_z$ are the Pauli matrices; $I$ is the identity matrix; $(0, \pm q_y, q_z)$ are the coordinates of the two Weyl points, which are symmetric with respect to the $x - z$ plane; and $\gamma < 1$ is a dimensionless parameter. The eigenvalues of the Hamiltonian are given by:



$$E(\boldsymbol{k}) = v_z(k_z - q_z) \pm \sqrt{v_x^2 k_x^2 + v_y^2(k_y^2 - q_y^2)^2 + v_z^2(k_z - q_z)^2 \gamma^2}$$

For a straightforward demonstration of Weyl point dispersion, we simply set the parameters as follows: $v_x = v_y = v_z = 1$, $q_y = q_z = 0.5$, and $\gamma = 0.5$, then the pair of Weyl point appear at $(0, \pm 0.5, 0.5)$. To illustrate the dispersion characteristics of the Weyl points, which spread linearly in all directions in 3D momentum space, Figs. S1 (a) and (b) show the dispersion of a pair of type-II Weyl points described by the effective Hamiltonian at fixed $k_z = 0.5$ and $k_x = 0$, respectively. The dispersion illustrates the anisotropy of type-II Weyl points, with both bands exhibiting positive group velocities in the $k_z$ direction.

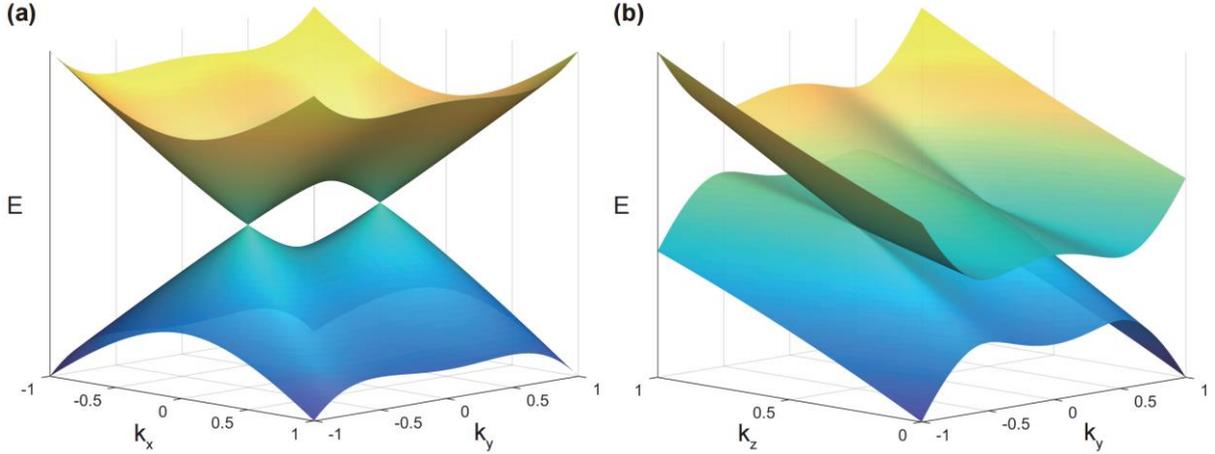

**Figure S1. Dispersion of a pair of type-II Weyl points under the effective Hamiltonian model: (a)** when $\boldsymbol{k_z = 0.5,}$ and **(b)** when $\boldsymbol{k_x = 0.}$

**III. The Berry phase evolution along the closed path in the (100) surface BZ.** The bulk bands corresponding to the closed yellow rectangular path (A-B-C-D-A) in the main text form a closed 2D sub-band system of the 3D structure. For a closed 2D photonic band system, the characteristics of bands and band gaps can be described by the concept of a well-defined Chern number. Employing the Wilson-loop method once more, we computed the Berry phase evolution along this path for the 1st, 2nd, and 3rd bulk bands, which represent the three bands



below the Weyl bandgap. As illustrated in Fig. S2, the Berry phase of the third band (indicated by the green dotted line) undergoes a change of $-2\pi$ (corresponding to a Chern number $C = -1$) around the closed yellow path, while the Berry phase of the first and second bands (indicated by the black dotted lines) remains unchanged. Hence, the total Chern number of the Weyl bandgap within this subsystem is $C = -1$, signifying one topological surface state across the Weyl bandgap on each (100) surface as illustrated in Fig. 3(b) of the main text.

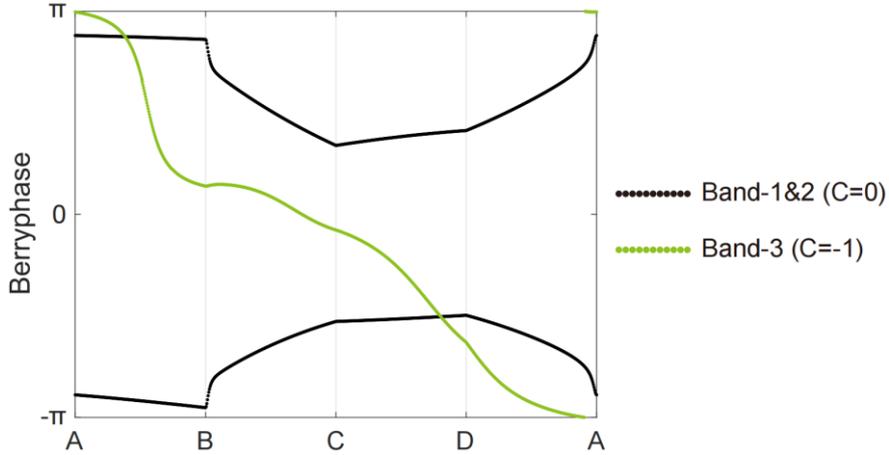

**Figure S2. The Berry phase evolution of the three bulk bands below the bandgap along the closed path depicted in Fig. 3(b) of the main text.**

**IV. The performance of the one-way fiber modes.** As commonly acknowledged, PCFs offer significant advantages in controlling crucial transmission parameters compared to conventional waveguides or fibers. In our analysis, we evaluate its potential performance by computing key parameters such as confinement loss [$L = \frac{20}{\ln(10)} \cdot \frac{2\pi}{\lambda} \cdot \text{Im}(n_{eff})$], group velocity ($v_g = d\omega/dk$, normalized by $c$, the speed of light in vacuum), and the dispersion parameter [$D_\lambda = \frac{\partial}{\partial \lambda} \frac{1}{v_g}$]. Here, $n_{eff}$ denotes the effective refractive index, $\lambda$ represents the wavelength, and $\beta$ signifies the wave vector. These parameters for the two Weyl guided modes are illustrated by the corresponding red and blue curves in Figs. S3(a-c).



The considered band range covers the extension range ($\Delta f$) of Mode-2 within the bandgap. As can be seen that the two topological fiber modes demonstrate a relatively low group velocity within the air-dominated hollow-core structure of the fiber, averaging about 0.3 times the speed of light. Moreover, they have a low dispersion parameter, with a zero-dispersion wavelength occurring at approximately 21 mm.

Furthermore, we conduct a separate analysis of Mode-2 (blue curves) to examine the relationship between confinement loss and the number of fiber layers (layers in the *x* direction multiplied by layers in the *y* direction), as depicted in Fig. S3(d). The findings clearly demonstrate that augmenting the number of cladding layers results in a substantial and swift decrease in confinement loss. They provide valuable insights into determining the optimal structural size required to meet specific transmission distance requirements.

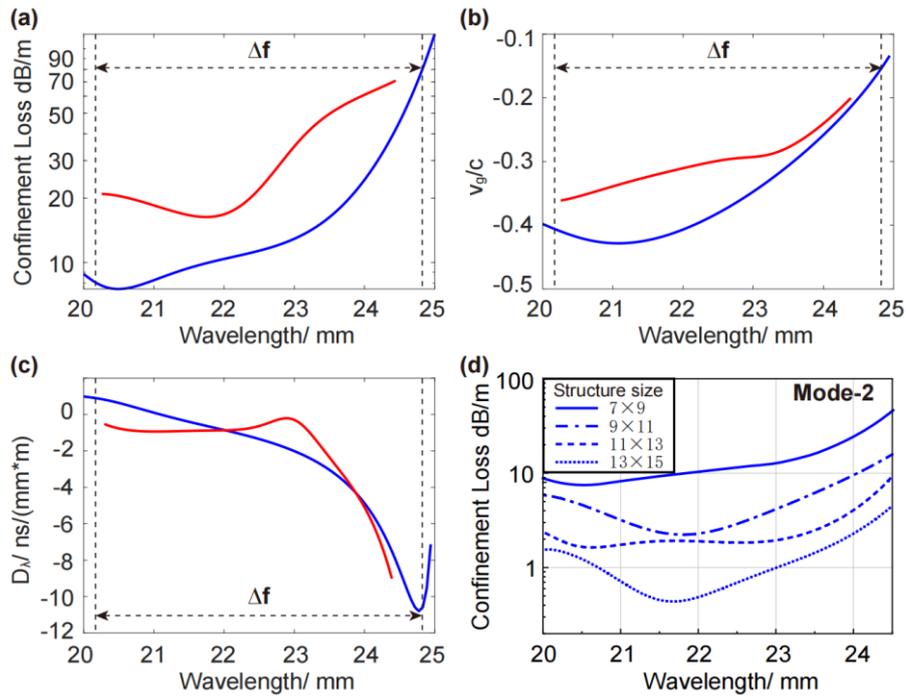

**Figure S3. The basic parameters of one-way fiber modes. (a-c)** Confinement loss, group velocity, and dispersion parameter of the two fiber modes (in 7 × 9 structure). **(d)** The confinement loss associated with Mode-2's dispersion curve varies across fiber structures with differing layer configurations.